\documentclass[prl,twocolumn,superscriptaddress,preprintnumbers,amsmath,amssymb]{revtex4}

\usepackage{graphicx}
\usepackage{dcolumn}
\usepackage{bm}

\begin{document}

\title{Comment on "Dispersive Terahertz Gain of a Nonclassical Oscillator: 
Bloch Oscillation in Semiconductor Superlattices"}

\author{A. Lisauskas}
\affiliation{Physikalisches Institut, Johann~Wolfgang~Goethe-Universit\"at,
D-60438 Frankfurt am Main, Germany}
\author{N. V. Demarina}
\affiliation{Electronics Department, Radiophysics Faculty, 
Nizhny Novgorod State University,
Nizhny Novgorod 603950, Russia}
\author{E. Mohler}
\author{H. G. Roskos}
\affiliation{Physikalisches Institut, Johann~Wolfgang~Goethe-Universit\"at,
D-60438 Frankfurt am Main, Germany}

\maketitle

In two recent Letters \cite{Hirakawa1,Hirakawa2}, Hirakawa et al. 
investigate the emission of coherent terahertz (THz) radiation from 
optically excited superlattices \cite{Waschke93}. The authors interpret 
the spectra of the THz transients in terms of a semiclassical transport 
model and conclude that the data provide evidence for the existence of 
the so-called Bloch gain which is of strong current interest because 
of its potential for the development of novel room-temperature THz 
sources \cite{Allen,lis}. The interpretation of the data in terms of a 
conductivity is, however, not performed correctly, thus making the 
conclusions of Refs.\,\onlinecite{Hirakawa1,Hirakawa2} void.

In Refs.\,\cite{Hirakawa1,Hirakawa2}, the authors employ an elegant 
way to analyze the response of the optically excited electrons to the 
constant bias field in the semiclassical approximation. They invoke 
the identity of this response with that of electrons present in a 
superlattice when the electric bias field is switched on abruptly 
($F(t) = F \Theta(t)$). Based on this analogy, a relationship between 
a time-dependent conductivity $\sigma(t)$ and the THz-field 
transient is derived ($\sigma(t) \propto E_{THz}(t) / F$, Eq.\,(6) 
in Ref.\,\cite{Hirakawa2}), which allows to relate the spectra of the 
THz transients directly to the frequency-dependent conductivity 
$\sigma(\omega)$.

The authors then make the mistake to interpret these data on 
the basis of an identification of $\sigma(\omega)$ with 
the theoretically derived function $\sigma_s(\omega)$ of 
Ref.~\onlinecite{ktitorov} which describes the small-signal 
response of electrons in a dc-biased superlattice to an 
additional ac electric field. Because of the highly nonlinear 
field dependence of the electron current, this response is 
generally very different from that to the switching 
of the full bias field as described by $\sigma(\omega)$. 

We illustrate the fundamental difference in Fig.~1, which 
displays theoretical results for the real parts of 
$\sigma_s(\omega)$ and $\sigma(\omega)$. Both terms can 
be expressed analytically, with $\sigma_{s}(\omega)$ given 
by \cite{ktitorov}
\begin{equation*}
 \sigma_{s}(\omega) = 
 \frac{\sigma_0}{1+\omega_B^2\tau_e\tau_p}\cdot\frac{1 
 -\omega_B^2\tau_e\tau_p -i\omega\tau_e}{(\omega_B^2-\omega^2)\tau_e\tau_p +1 
 -i\omega(\tau_e+\tau_p)}, 
\end{equation*}
and $\sigma(\omega)$ obtained by Fourier transformation of $\dot{v}(t)$ of \cite{Ignatov91}
\begin{equation*}
 \sigma(\omega) = \sigma_0\frac{1 - 
 i\omega\tau_e}{(\omega_B^2-\omega^2)\tau_e\tau_p +1 
 -i\omega(\tau_e+\tau_p)}.
\end{equation*}
Here, $\sigma_0$ is the static low-field conductivity, $\omega_B 
\propto F$ the Bloch frequency, and $\tau_e$, $\tau_p$ the energy 
and momentum relaxation times (values equal to those of 
Ref.\,\onlinecite{Hirakawa2}).

\begin{figure}[!ht]
	\includegraphics[width=7cm]{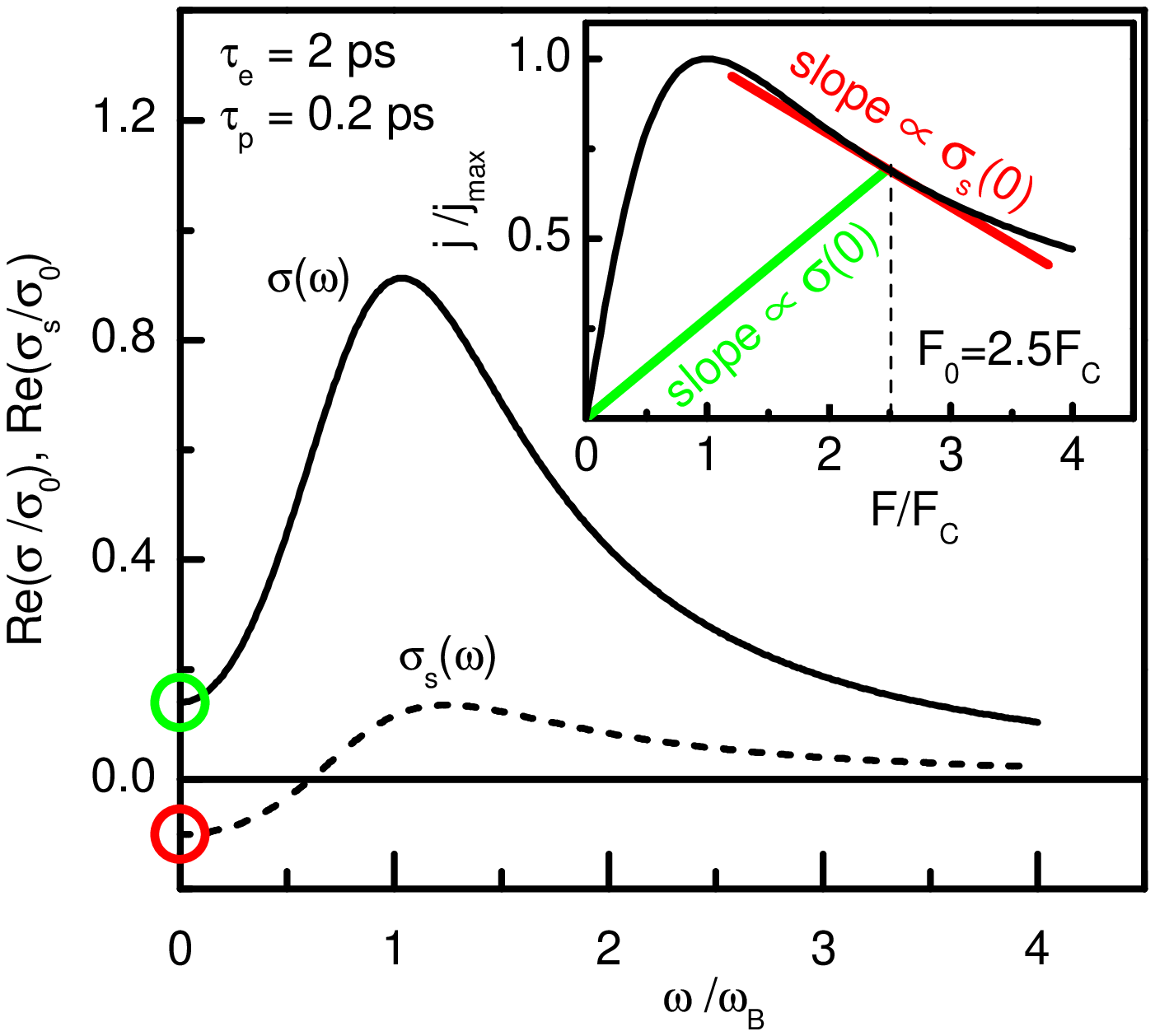}
	\label{fig1}
	\caption{Main panel: Real part of conductivity functions $\sigma(\omega)$ 
	(full line) and $\sigma_s(\omega)$ (dashed line). Inset: 
	current density in a superlattice vs. electric field, and 
	illustration of the different nature of the conductivities for 
	$\omega \rightarrow 0$. While $\sigma_s(0)=dj/dF|_{F_0}$, the 
	small-signal gain at a chosen bias field $F=F_0$, is given by 
	the slope of the $j(F)$ curve, the full-field-switching 
	response $\sigma(0)$ is given by the slope $j/F_0$.} 
\end{figure}

Bloch gain is expected for electric fields exceeding the 
critical value $F_C$ (defined by the maximum of the current/field 
curve shown in the inset of Fig.\,1). Indeed, Re($\sigma_{s}(\omega)$) 
is negative for $\omega < 1/\tau_e \sqrt{(\omega_B^4\tau_e^2\tau_p^2-1) 
/ (\omega_B^2\tau_p^2+1)}$ which is indicative for gain at these 
frequencies. In contrast, Re($\sigma(\omega)$) is positive for 
all frequencies and does not evince a gain signature. 

These results show that one must not compare spectra 
of THz transients of the kind discussed here with the small-signal 
response function $\sigma_s(\omega)$. We finally note, that the 
negative Re($\sigma(\omega)$) values derived from the experiments 
in Ref.\,\onlinecite{Hirakawa2} either have a different 
(non-semiclassical) origin or are artifacts of the 
interpolation of the zero-time-delay position.


\end{document}